\documentclass{webofc}
\usepackage[varg]{txfonts}   
\usepackage{multirow} 
\usepackage{amsmath} 
\usepackage{amssymb}
\usepackage{epsfig} 
\usepackage{graphicx} 
\usepackage{color}
\usepackage{xspace} 
\usepackage{lineno}
\usepackage[varg]{txfonts}   
\usepackage{subfigure}
\newcommand{\dAu}{\mbox{${\rm {\it d}+Au}$}\xspace}

\newcommand{\pAl}{\mbox{${\rm {\it p}+Al}$}\xspace}

\newcommand{\HeAu}{\mbox{${\rm ^{3}He+Au}$}\xspace}

\newcommand{\UU}{\mbox{${\rm U+U}$}\xspace}

\newcommand{\CuAu}{\mbox{${\rm Cu+Au}$}\xspace}
\newcommand{\pp}{\mbox{${p+p}$}\xspace}

\newcommand{\raa}{\mbox{$R_{AA}$}\xspace}

\newcommand{\snn}{\mbox{$\sqrt{s_{_{NN} }}$}\xspace}

\newcommand{\pt}{\mbox{$p_{T}$}\xspace}

\usepackage{xspace}	

\newcommand{\mt}{\mbox{$m_T$}\xspace}
\newcommand{\ut}{\mbox{$\left<u_t\right>$}\xspace}
\newcommand{\To}{\mbox{$T_0$}\xspace}

\newcommand{\rab}{\mbox{$R_{AB}$}\xspace}
\newcommand{\Npart}{\mbox{$\langle N_{\rm part} \rangle$}\xspace}

\newcommand{\pal}{\mbox{$p$$+$Al}\xspace}
\newcommand{\heau}{\mbox{$^3$He$+$Au}\xspace}

\newcommand{\cuau}{\mbox{Cu$+$Au}\xspace}
\newcommand{\uu}{\mbox{U$+$U}\xspace}

\newcommand{\aprot}{\mbox{$\bar{p}$}\xspace}
\newcommand{\prot}{\mbox{$p$}\xspace}
\newcommand{\prots}{\mbox{$(p+\bar{p})/2$}\xspace}
\newcommand{\Km}{\mbox{$K^-$}\xspace}
\newcommand{\Kp}{\mbox{$K^+$}\xspace}
\newcommand{\Kpm}{\mbox{$K^\pm$}\xspace}
\newcommand{\pim}{\mbox{$\pi^-$}\xspace}
\newcommand{\pip}{\mbox{$\pi^+$}\xspace}
\newcommand{\pipm}{\mbox{$\pi^\pm$}\xspace}

\def\FigureOne{
\begin{figure*}[!]
\centering
\includegraphics[width=0.9\linewidth]{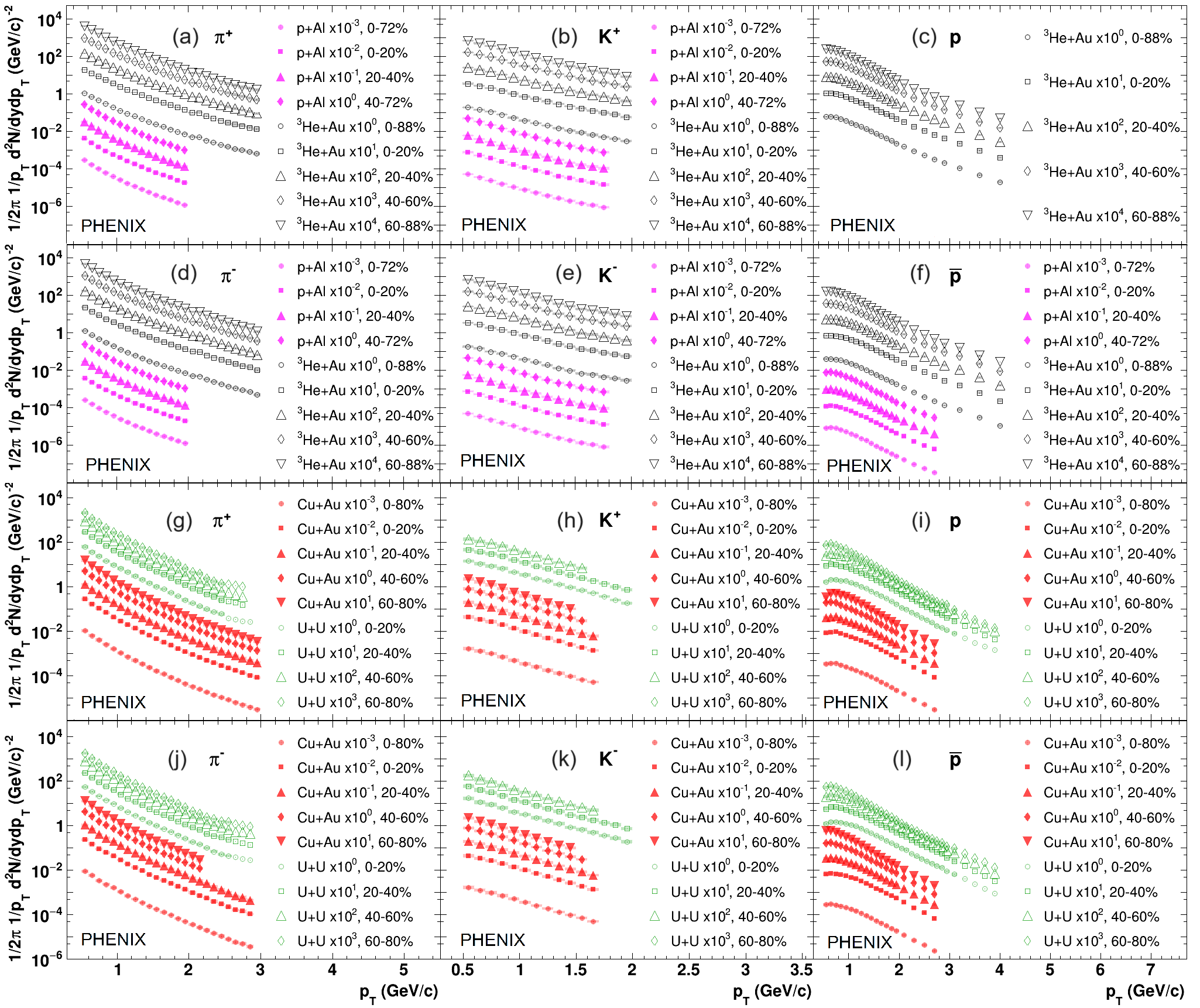}
\vskip -0.2cm
\caption{The \pip, \pim, \Kp,\Km, \prot and \aprot invariant \pt spectra 
measured in different centralities of \pAl, \HeAu, \CuAu collisions at 
\snn = 200 GeV and \UU collisions at \snn = 193 GeV.}
\label{fig:fig1}
\end{figure*}
}

\def\FigureTwo{
\begin{figure*}[!bht]
\vskip -0.2cm	
\subfigure[Mass and centrality dependence of inverse slope]
{
	\includegraphics[width=0.48\linewidth]{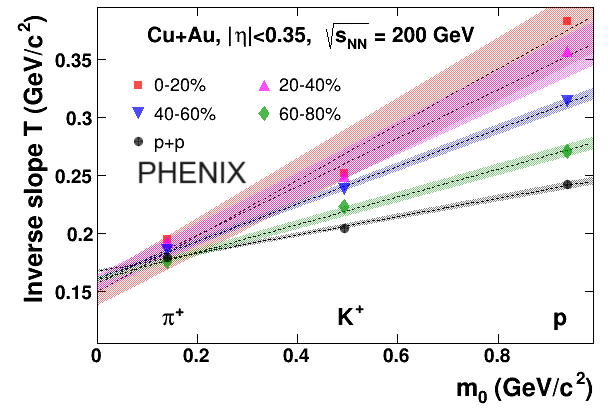}
\label{fig:Tinv}	
}
\subfigure[Freeze-out temperature as a function of \Npart]
{
	\includegraphics[width=0.45\linewidth]{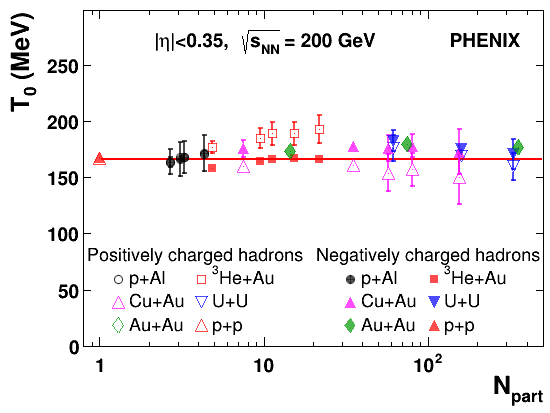}
\label{fig:T0Npart}
}
\vspace*{-0.3cm}	
\caption{(a) Mass and centrality dependence of inverse slope parameters $T$ 
for \pip, \Kp and \prot in Cu$+$Au collisions. The dotted lines represent a linear fits (see text). (b) Freeze-out temperature (\To) as a function of \Npart obtained for positively and negatively charged hadrons at different centralities of collision systems. The \To values measured in \pp collisions are shown for comparison.}
\label{fig:fig2}
\end{figure*}
}
\def\FigureThree{
\begin{figure*}[!bht]
\centering
\vskip -1cm
\includegraphics[width=0.8\linewidth]{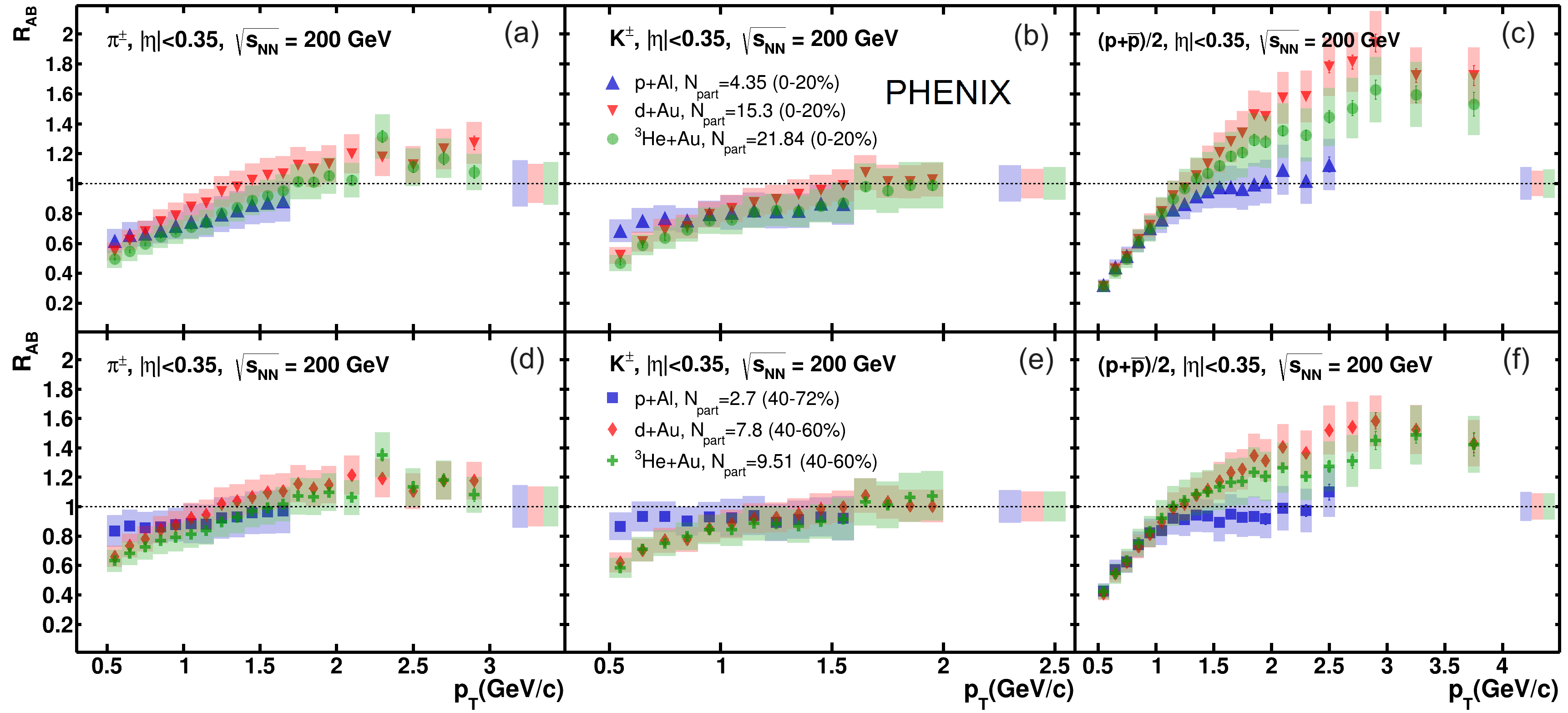}
\vskip -0.2cm
\caption{Identified charged hadron nuclear-modification factors as a 
function of \pt measured in central and peripheral \pal, \dAu and \heau 
collisions. The dashed lines correspond to \rab = 1 indicating absence of nuclear modification.  }
\label{fig:RAB_small}		
\end{figure*}
}
\def\FigureFour{
\begin{figure*}[!hbt]
	\subfigure[\raa for central and peripheral \CuAu, and \UU]
		{
	\includegraphics[width=0.5\linewidth]{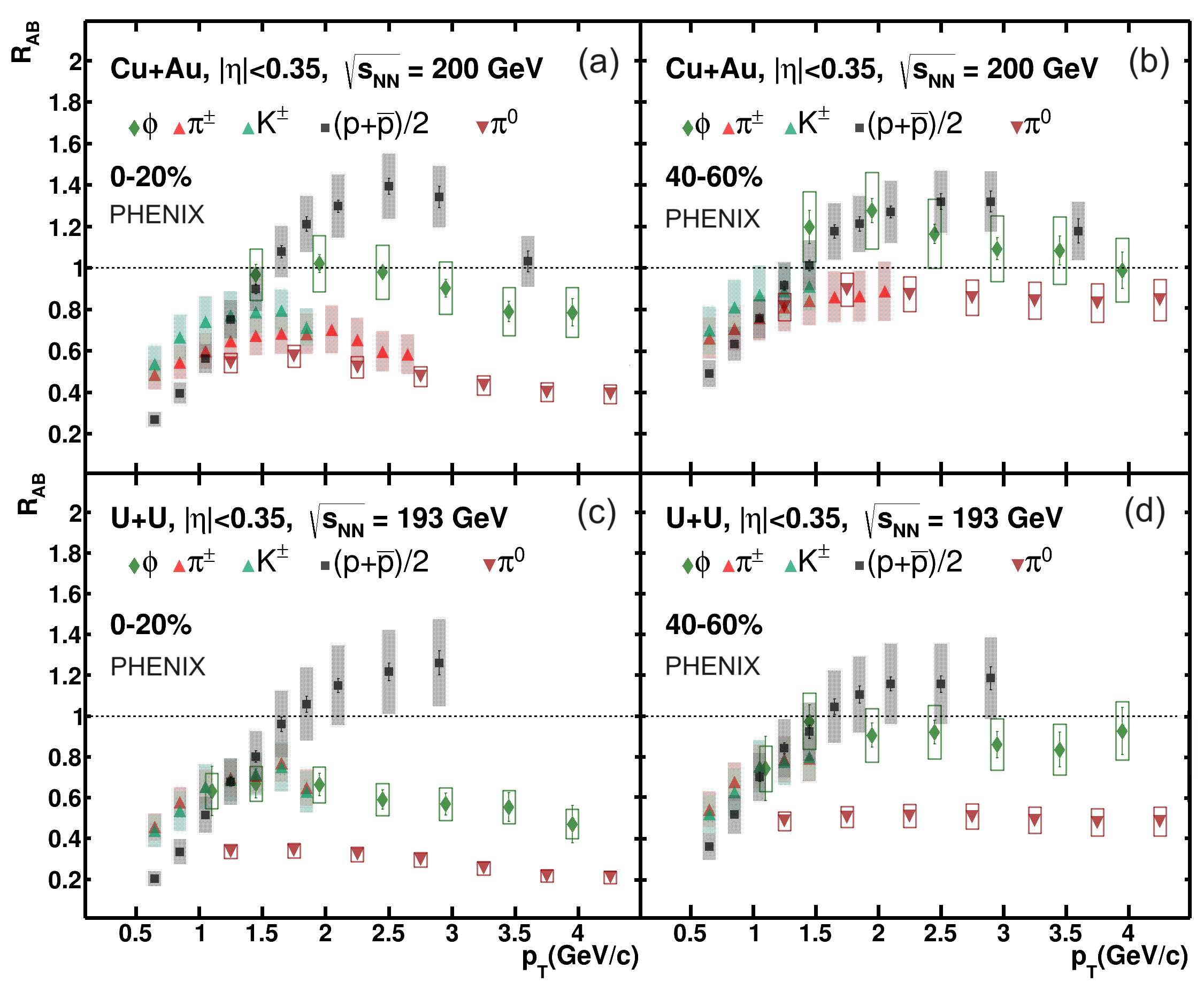}
	\label{fig:RabMesons_large}						
		}
	\subfigure[\raa for central and peripheral \pAl, and \heau]
		{
	\includegraphics[width=0.5\linewidth]{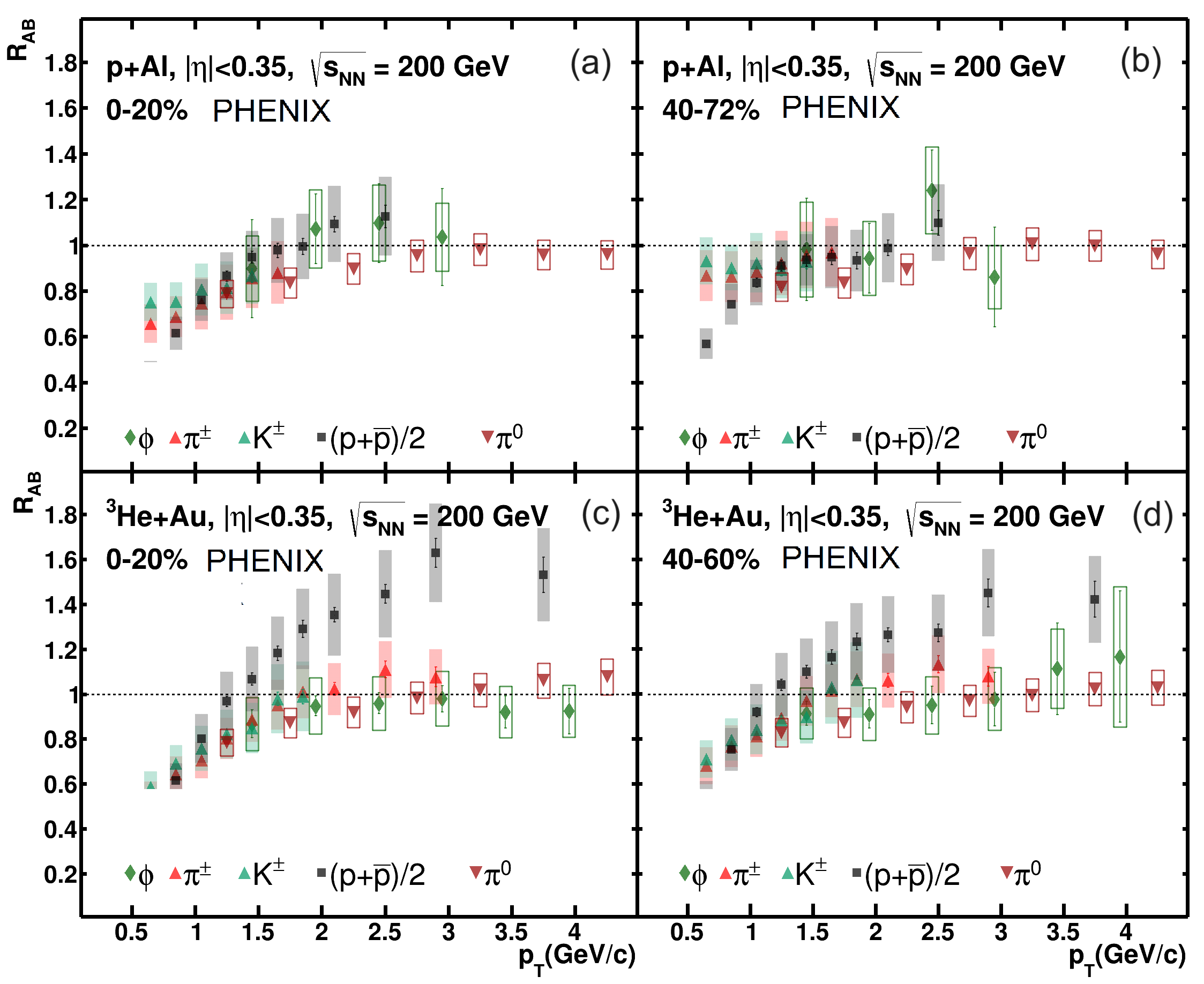}
	\label{fig:RabMesons_small}	
		}
		\vskip -0.3cm
\caption{\rab for light hadron, $\phi$, \pipm, \Kpm, \prots and $\pi^0$, as a function of \pt measured in central and peripheral: (a) in Cu$+$Au, and U$+$U collisions, and (b) $p$$+A$l and \heau collisions.}	
\label{fig:fig4}	
\end{figure*}
}
\begin{document}

\title{Scaling Properties of ${\LARGE \phi}$-Meson and Light Charged
  Hadron Production in Small and Large Systems at PHENIX}
\makeatother  
\author{\firstname{Rachid} \lastname{Nouicer}
\inst{1}\fnsep\thanks{\email{nouicer@bnl.gov}}\ (for the
  PHENIX Collaboration) } \institute{Physics Department, Brookhaven
  National Laboratory, Upton, New York 11973, United States}

\abstract{
Recent results on the identified charged-hadron ($\pi^\pm$, $K^\pm$, $p$, $\bar{p}$) production at midrapidity region ($|\eta|<$ 0.35) have been measured by the PHENIX experiment in \pAl, \HeAu, \CuAu collisions at \snn~=~200~GeV and \UU collisions at \snn~=~193~GeV. These measurements are presented through the invariant transverse-momentum ($p_T$) and transverse-mass ($m_T$) spectra for different collision centralities. The averaged freeze-out temperature value for different systems was found to be \mbox{$166.1 \pm 2.2$ MeV}, and do not exhibit any dependence on the collision centrality and \Npart values. The particle ratios of $K/\pi$ and $p/\pi$ have been measured in different centrality ranges of large and small  collision systems. The values of $K/\pi$ ratios measured in all considered collision systems were found to be consistent with those measured in $p$$+$$p$ collisions. Furthermore, the identified charged-hadron nuclear-modification factors ($R_{AB}$) are also presented. Enhancement of proton $R_{AB}$ values over meson $R_{AB}$ values was observed in central $^3$He$+$Au, Cu$+$Au, and U$+$U collisions. The proton $R_{AB}$ values measured in $p$$+$Al collision system were found to be consistent with $R_{AB}$ values of $\phi$, $\pi^\pm$, $K^\pm$, and $\pi^0$ mesons, suggesting that the size of the system produced in $p$$+$Al collisions is too small for recombination to cause a noticeable increase in proton production.
}
\maketitle
\vspace*{- 0.5cm}
\section{Introduction \label{sec-1}}
Light hadrons are considerably produced in high-energy heavy-ion collisions and provide a wealth of information about properties of created QCD medium and reaction dynamics. These include, in particular, the implications of collective flow in small and large systems and the impact of recombination on baryon and strangeness enhancement. The system size dependence studies of different observable are crucial to investigate the properties of quark-gluon plasma (QGP) and hadronization based on initial conditions of the collisions. 
   
This paper presents recently finalized measurements by
PHENIX on the production of identified charged hadron ($\pi^\pm$, $K^\pm$, $p$, $\bar{p}$) in small and large system size, $p$+Al, $p$/d/${^3}$He/Cu+Au collisions at \snn = 200 GeV, and U+U collisions at \snn = 193 GeV, and results were published recently~\cite{ppg249}. The data sets used in the present analysis were collected by the PHENIX experiment~\cite{PHENIXoverview}. The latter consisted of three components: global detectors, the central spectrometers, and the muon arm spectrometers. These measurements use the central spectrometers ($|\eta|<0.35$), and provide precise tracking and particle identification for electrons, charged hadrons, and photons. Tracking and momentum determination are provided by drift chambers and pad chambers. Particle identification is provided by time of flight, ring-imaging Cherenkov detectors, and  electromagnetic calorimeters.
\section{Identified Charged Hadron Results in Small and Large Systems\label{sec-2}}
   \FigureOne
{\textbf{ $\bullet$ Invariant Spectra and Freeze-out Temperature}}: figure~\ref{fig:fig1} shows the \pipm, \Kpm, \prot, and \aprot invariant transverse momentum spectra measured in \pal, \heau, \cuau collisions at \snn = 200 GeV 
and in \UU collisions at \snn~=~193 GeV. We observe that the $\pi$, $K$, and $p$ invariant spectra reveal different shapes as a function of \pt. To assess these differences, invariant-transverse-mass ($m_T=\sqrt{p_T^2 + m_0^2}$) spectra were calculated. We observe that the \mt invariant spectra of all identified charged hadrons have exponential form for \mt $<$ 1.5~GeV and was approximated using following fit function:
\begin{equation}
\label{eq:spectra_mt}
\frac{1}{2\pi m_T} \frac{d^2 N}{dm_T dy}= 
\frac{A  }{2\pi T (T+m_0)}\exp \left( -\frac{m_T - m_0}{T}\right)
\end{equation}
where $T$ is the inverse-slope parameter, $A$ is a 
normalization factor, and ${m_0}$ mass of the charged particle at rest. Figure~\ref{fig:Tinv} shows examples of $T$ parameter vs.~hadron mass 
($m_0$) dependencies for different centralities of \CuAu collisions. 
Ordering of pion, kaon, and proton inverse slope values  $T_{\pi}<T_{K}<T_p$ can be seen in all centralities. By doing similar work in all systems, \pal, \heau, and \UU, we obtained that the $T$ values, calculated for pions in different centralities, are nearly of the same values in all collision systems. The $T$ values, calculated for kaons, take intermediate values between pion and proton $T$-parameter values. The dotted lines on Fig.~\ref{fig:Tinv} represent linear fits of the $T(m_0)$ values from each centrality bin using thermal formula
$\label{eq:InvSlope} 
T = T_0 +m \left< u_t\right>^2,$ 
where $T_0$ can be interpreted as a freeze-out temperature and \ut as the average collective velocity for all particle species. The fit parameters for positively charged ($T_{0}^{+}$, 
$\left<u_{t}\right>^{+}$) and negatively charged ($T_{0}^{-}$, 
$\left<u_{t}\right>^{-}$) hadrons calculated in \pal, \heau, Cu$+$Au, and 
U$+$U collision systems are presented in Fig.~\ref{fig:T0Npart} as a function of \Npart values. The $T_0$ values calculated in collisions with different geometries and centralities were found to be coincident within uncertainties, indicating that the freeze-out temperature is approximately independent of \Npart values. The averaged $T_0$ value was found to be \mbox{$166.1 \pm 2.2$ MeV} and is shown in the Fig.~\ref{fig:T0Npart} with red solid line. 
\FigureTwo
\FigureThree

\noindent {\textbf{ $\bullet$ Particle Ratios:}}
the baryon production enhancement in nucleus-nucleus collisions is contemplated to be one of the signatures of QGP formation~\cite{ppg026}. Based on this consideration, the ratios of \prot/\pip, \aprot/\pim in different centralities of large (\cuau, \uu) and small (\pp, \pal, \heau) collision systems have been measured and published recently by the PHENIX collaboration~\cite{ppg256}. We observed in central collisions of large systems the $p$/$\pi$ ratios reach the values of $\approx$ 0.6, but in peripheral collisions the values of $p/\pi$ ratios are smaller than 0.4 in the whole \pt range. Furthermore, in small collision systems (\pal, \heau), the values of $p$/$\pi$ ratios are in good agreement to those measured in \pp collisions~\cite{CH_pp}.
\FigureFour
\vskip 0.2cm 
\noindent{\textbf{ $\bullet$ Nuclear Modification Factor:}}
figure~\ref{fig:RAB_small} shows the identified charged-hadron \rab values as a function of \pt in central and peripheral \pal, \dAu, and \heau collisions. The \rab values 
are found similar for collision systems with different geometries, but with the same \Npart values, indicating that identified charged-hadron production depends only on system size and not geometry. In addition, we observe the following features: 1) the slope of $R_{AB}(p_T)$ in \pal collisions is flatter than it is in \heau and \dAu collisions, and 2) proton \rab values in \pal collisions at the intermediate \pt range (1.0~GeV/$c$~$<~p_T~<$~2.5 GeV/$c$) are equal to unity, while in \heau and \dAu collisions proton \rab is above unity.
These observations suggest that the differences between \pal and $d/$\heau might be caused by the size of the \pal system being insufficient to observe an increase in proton production.
\vskip -0.3cm
\section{Comparison of Identified Charged  Hadron with ${\LARGE \phi}$-Meson and $\pi^0$ \label{sec-3}}
Figures~~\ref{fig:RabMesons_large} and \ref{fig:RabMesons_small} show the comparison of \rab values  of identified charged-hadron with ${\LARGE \phi}$-meson and $\pi^0$ in small and large collision systems, respectively. 
All these measurements of \rab were obtained by the PHENIX experiment~\cite{ppg256,phi}. 
These results elucidate following features: 1) in large collision systems and in the $^{3}$He+Au collision system, proton \rab values 
are enhanced over all meson \rab values. Knowing that the mass of the $\phi$-meson \mbox{$m_{\phi}$=1019 MeV/$c^2$} is similar to the proton mass \mbox{$m_{p}$=938 
MeV/$c^2$}, therefore the enhancement of proton $R_{AB}$ values over $\phi$-meson $R_{AB}$ values suggests differences in baryon versus meson production instead of a simple mass dependence, 2) in \pal collisions 
proton $R_{AB}$ values and $R_{AB}$ values of all measured mesons are in good agreement within uncertainties.

\section{Summary \label{sec-4}} 
The PHENIX experiment has established a comprehensive program to study the light hadron  ($\phi$, \pipm, \Kpm, \prots and $\pi^0$) production in small and large collisions systems by measuring invariant transverse-momentum and transverse-mass spectra, particle ratios, and nuclear modification factors. These results were presented and discussed in the scope to study the properties of created QCD medium and reaction dynamics at RHIC energies.

\end{document}